\documentstyle[aps,10pt]{revtex}

\begin{document}

\title{Analytic estimation of Lyapunov exponent in a\\
mean-field model undergoing a phase transition}
\author{Marie-Christine Firpo\thanks{%
firpo@newsup.univ-mrs.fr}}
\address{Equipe turbulence plasma de l'UMR 6633 \\
CNRS--Universit\'e de Provence\thanks{Address from October 1997: 
Case 321,
Centre de Saint-J\'er\^ome, Av. Escadrille Normandie Niemen,
F-13397 Marseille Cedex 20}, \\
IMT Ch\^ateau-Gombert, F-13451 Marseille cedex 20}
\date{\today}
\maketitle

\begin{abstract}
\parbox{14cm}

The parametric instability contribution to the largest Lyapunov
exponent $\lambda _{1}$ is derived for a mean-field Hamiltonian model, with 
attractive long-range interactions. This uses a recent Riemannian 
approach to describe Hamiltonian chaos with a large number $N$ of degrees of freedom.
Through microcanonical estimates of suitable geometrical 
observables, the mean-field behavior of $\lambda _{1}$ is 
analytically computed and related to the second order phase 
transition undergone by the system. It predicts that chaoticity 
drops to zero at the critical temperature and remains vanishing above it, 
with $\lambda _{1}$ scaling as $N^{-\frac{1}{3}}$ to the leading order in $N$.

\end{abstract}

%\pacs{05.45.+b,05.70.Fh,02.40.-k} 

\vskip 0.5cm


\noindent 05.70.Fh (Phase transitions: general aspects)

\noindent 02.40.-k (Geometry, differential geometry, and topology)


%\noindent preprint TP 97.08 
%{\it private communication / to be submitted for publication}

% Introduction

\section{Introduction}

The largest Lyapunov exponent $\lambda _{1}$ is a good
quantity to measure the degree of chaoticity of a
generic non-integrable Hamiltonian system.
However its numerical computation requires
to compute also the microscopic dynamics for, sometimes, very long and,theoretically,
 infinite time. This
may obviously turn rapidly difficult to tackle and much effort has been
devoted to derive some asymptotic scaling laws \cite{Paladin86} and, more
recently, to get analytic estimates by relating microscopic dynamics with
statistical averages, provided the number $N$ of degrees of freedom is large 
enough \cite{Casetti95,Casetti96,Caiani97}. 
This latter way of analytically computing $%
\lambda _{1}$ as a function of $\varepsilon =E/N$, the energy per degree of
freedom, has proved to be remarkably efficient. It reformulates Hamiltonian
dynamics in the language of Riemannian geometry, using the fact that the natural
motions can be viewed as geodesics of a suitable Riemannian manifold \cite{Pettini}.
Chaotic motion then reflects into the instability of the geodesic flow which
depends on curvature properties of the manifold. This geometric
formulation of the dynamics has been known for long and led to fundamental results
in abstract ergodic theory when the ergodicity of geodesic flows on compact
manifolds of negative curvature was demonstrated by Hedlund and Hopf in
1939, and later exploited by Krylov \cite{Szasz}. However, when more
physical Hamiltonian systems come into play, such as coupled nonlinear
oscillators, a major source of chaos appears to be parametric
instability activated by a fluctuating curvature along the geodesics, even
when curvature is always positive \cite{Cerruti,Kandrup}%
. This has been exploited in the theoretical model proposed by M. Pettini and 
coworkers. Modeling the effective curvature felt by a geodesic by a
gaussian stochastic process, with mean the average Ricci curvature and variance
its fluctuations, and under the ergodic hypothesis replacing the previous
geometrical quantities with their averages $\kappa _{0}$ and $\sigma
_{\kappa }^{2}$ according to the natural ergodic measure, i.e. in the
microcanonical ensemble, they derive the following expression for $\lambda
_{1}$ \cite{Casetti95,Casetti96}:

\begin{equation}
\lambda _{1}=\frac{\Lambda }{2}-\frac{2\kappa _{0}}{3\Lambda }
\label{lambda}
\end{equation}
with: 
\begin{equation}
\Lambda =\left( 2\sigma _{\kappa }^{2}\tau +\sqrt{\frac{64}{27}\kappa
_{0}^{3}+4\sigma _{\kappa }^{4}\tau ^{2}}\right) ^{\frac{1}{3}}
\label{glambda}
\end{equation}
and $\tau$, a timescale for the stochastic process estimated as:
\begin{equation}
\tau =\frac{%
\pi \sqrt{\kappa _{0}}}{2\sqrt{\kappa _{0}}\sqrt{\kappa _{0}+\sigma 
_{\kappa }}%
+\pi \sigma _{\kappa }}
\label{tau}
\end{equation}

In this article, we apply these geometrical tools to a mean-field
Hamiltonian system of globally coupled rotators exhibiting a second order
phase transition at a certain critical energy $\varepsilon _{c}$. We
analytically estimate the parametric instability contribution to $\lambda
_{1}(\varepsilon )$ and predict a neat distinction between the two cases: $%
\varepsilon <\varepsilon _{c}$ and $\varepsilon >\varepsilon _{c}$.
Numerical simulations \cite{Latora97,Yama96} seem to qualitatively support
the analytical conclusions. The remarkable behavior of the Lyapunov
exponent in the mean-field limit, as a consequence of the simple expressions
of relevant geometrical quantities as functions of the order parameter,
could then be a dynamical signature of the phase transition.\\
The model at 
hand will be described in Sec. II and some useful geometric expressions 
derived there. A detailed derivation of the largest Lyapunov 
exponent $\lambda_{1}$ as a function of the energy density $\varepsilon$ 
will be exposed in Sec. III, Sec. IV being devoted to comments and 
conclusions.

% The model

\section{The mean-field model and first useful geometric expressions}

Here we study the so-called mean-field Hamiltonian X-Y model,
which can be considered as a toy-model for investigating long-range
interactions in Coulomb systems \cite{Antoni95,Ruffo94} . The dynamics of $N$
interacting particles moving on the unit circle $\Pi=[0;2 \pi]$ derives 
from the following Hamiltonian 
\begin{equation}
H=\sum_{l=1}^N {\frac{p_{l}^{2}}{2}}+\frac{c}{2N}%
\sum_{l,r=1}^N {\left[ 1-\cos (q_{l}-q_{r})\right]}
=K+V(q)
\label{H}
\end{equation}
where $K$ and $V$ stand for the kinetic and the potential energy
respectively. Constant $c$ may be rescaled to $+1$, $0$ or $-1$ 
by a change of variables. 
The scaling factor $\frac{1}{N}$ for the potential energy
ensures that the interaction energy is extensive and emphasizes its
mean-field nature. Thus, in the following, we would not deal with the
usual thermodynamic limit with fixed density, but rather with the mean-field
limit $N\rightarrow \infty $, $\frac{H}{N}\rightarrow \varepsilon $, $%
\varepsilon $ finite. Note that the total momentum is also a constant 
of the motion. However this will not affect following calculation since 
the potential only depends on positions.

The equilibrium statistical mechanics of this model can be exactly derived 
\cite{Elskens97}. In the case of an attractive potential (i.e. $c>0$%
), which will be assumed in the following, that is in the ferromagnetic-like case, 
it predicts a second-order phase
transition with order parameter $\left\| {\bf M}\right\| $ where ${\bf M}$ is
the mean-field magnetization-like variable defined as:

\begin{equation}
{\bf M}=\left( \frac{1}{N} \sum_{l=1}^N {\cos (q_{l})},%
\frac{1}{N} \sum_{l=1}^N {\sin (q_{l})} \right)
\end{equation}
This phase transition can be easily conjectured by observing 
that at small energy $\left\| {\bf M}\right\| =O\left( 1\right) $ 
with a clustered phase, whereas at large
energy, the central limit theorem predicts that $\left\| {\bf M}\right\|
=O(N^{-\frac{1}{2}})$ with particles having random ballistic motions. It is
also interesting to note that introducing the global variable ${\bf M}$
enables to re-express the equation of motion of any particle as:

\begin{equation}
\ \ddot{q}_{i}=-c\left\| {\bf M}\right\| \sin (q_{i}-\phi )\text{ where }\phi
=\arg ({\bf M})
\label{pendule}
\end{equation}
that is the equation of a perturbed pendulum, -the full system being closed
by adding the evolution equations for $\left\| {\bf M}\right\| $ and $\phi $.

Let us now first express in the framework of the Eisenhart metric the Ricci
curvature associated to this system, then derive the microcanonical averages
of the geometrical quantities involved, via the canonical ensemble 
which leads to simpler calculations. Recall
here that in the limit of infinite size, that is $N\longrightarrow \infty $, 
the averages of thermodynamic observables in different 
ensembles coincide \cite{Anto}, but not their 
fluctuations \cite{Balian}. Therefore, in order to get the fluctuations of an observable
 $f$ in the 
microcanonical ensemble, it will be necessary to add a corrective term 
according to the formula derived in \cite{Lebo}, which is not valid 
at the critical point:

\begin{equation}
\left\langle \delta ^{2}f\right\rangle _{\mu }=\left\langle \delta
^{2}f\right\rangle _{c}+\left( \frac{\partial \left\langle \varepsilon
\right\rangle _{c} }{\partial \beta }\right) ^{-1}\left[ \frac{\partial
\left\langle f\right\rangle _{c}}{\partial \beta }\right] ^{2}
\label{correc}
\end{equation}

where \cite{note1},
 
\begin{equation}
\left\langle \delta ^{2}f\right\rangle \equiv
\frac{1}{N}\left\langle
\left( f-\left\langle f\right\rangle \right) ^{2}\right\rangle
\end{equation}

So, with the Eisenhart metric, Ricci curvature reads $K_{R}(q)=\Delta V$,
where $\Delta $ stands for the euclidian Laplace operator in the configuration 
space, so that the average Ricci curvature \cite{Casetti96}, defined 
as $k_{R}(q)\equiv \frac{K_{R}(q)}{N-1}$, is

\begin{equation}
k_{R}(q)=\frac{1}{N-1} \sum_{i=1}^N {\frac{\partial
^{2}V(q)}{\partial q_{i}^{2}}}=c-\frac{2}{N-1}V(q)
\end{equation}

Moreover, a straightforward calculation gives:

\begin{equation}
V(q)=\frac{cN}{2}(1-\left\| {\bf M}\right\| ^{2})
\end{equation}
Thus we obtain the key expression that the mean Ricci curvature reads
simply in terms of the order parameter, the mean-field magnetization ${\bf M}$ as:
\begin{equation}
k_{R}=c\left\| {\bf M}\right\| ^{2}
\label{courbem}
\end{equation}
up to a $O\left( N^{-1}\right)$ term, which, as far as the mean-field limit 
is concerned, gives a vanishing contribution and 
will be ignored. It will only play a part in corrections 
above the transition. It should be pointed out that this expression 
for the mean Ricci curvature as a smooth function of the natural order 
parameter, the magnetization, is not claimed here (since not proved) to be 
a generic property of, for instance, some class of mean-field Hamiltonian 
systems. At present we should thus consider the results obtained in this 
article as peculiar features of the model at hand.
As only positions-involving quantities come into play, let us now focus on
the contribution of the potential energy to the partition function in the
canonical ensemble at temperature $T=\beta^{-1}$ (with $k_{B}=1$):

$Z_{c}(\beta )= \int_{\Pi ^{N} }\exp (-\beta V(q))d^{N}q=\exp
(-\beta \frac{cN}{2}) \int_{\Pi ^{N} } \exp (\beta \frac{cN}{2}%
\left\|{\bf M}\right\|^{2})d^{N}q$

Then, using the integral representation of gaussian
functions, we get:

$Z_{c}(\beta )=\exp (-\beta \frac{cN}{2}) \int_{\Pi ^{N} } \frac{1%
}{\pi }\left[ \int_{{I\!\!R}^{2}} \exp (-{\bf u}^{2}+2\sqrt{\beta \frac{%
cN}{2}}{\bf u}.{\bf M})d{\bf u}\right] d^{N}q$ 

$=\exp (-\beta \frac{cN}{2})\frac{(2\pi )^{N}}{\pi } \int_{{I\!\!R}^{2}} %
\exp (-{\bf u}^{2}) \left[ I_{0}\left( 2\sqrt{\beta \frac{c}{2N}}\left\| {\bf u}%
\right\| \right)\right] ^{N}d{\bf u}$

$=(2\pi )^{N}\frac{N}{\beta c} \int_0^\infty%
 rdr\exp \left( -N\psi (r,\beta )\right) $

where $\psi (r,\beta )\equiv \frac{r^{2}}{2\beta c}-\ln (I_{0}(r))+\frac{%
\beta c}{2}$ and where $I_{n}$ stands for the modified Bessel 
function of order $n$.

Then, according to the saddle-point method, in the limit $N\rightarrow
\infty $ the previous integral is fully dominated by the minimum of $\psi $
obtained by solving the consistency equation 
$\partial _{r}\psi (r,\beta )=0$ that is:

\begin{equation}
\frac{r}{\beta c}-\frac{I_{1}(r)}{%
I_{0}(r)}=0  
\label{cons}
\end{equation}

When $\beta c<2$, $\psi $ is minimal for $r=0$, which corresponds to a
vanishing magnetization. For $\beta c>2$, (\ref{cons}) admits a non-vanishing
solution noted $r^{*}(\beta )$, the phase transition taking place for $\beta
c=2$ i.e. for $T_{c}=c/2$ and $\varepsilon_{c}=3c/4$.

Before examining these two cases, we establish some useful canonical
relations: as $\left\langle V(q)\right\rangle _{c}=-\partial _{%
\beta }\ln (Z_{c})$ and 
$\left\langle \left( V(q)-\left\langle
V(q)\right\rangle \right) ^{2}\right\rangle _{c}=\partial _{\beta }^{2}\ln
(Z_{c})$, one obtains respectively: 

\begin{equation}
\left\langle k_{R}\right\rangle _{c}=c+\frac{2}{N}\partial _{\beta }\ln
(Z_{c})
\end{equation}

\begin{equation}
\left\langle \delta ^{2}K_{R}\right\rangle _{c} \equiv \frac{1}{N}\left\langle
\left( K_{R}-\left\langle K_{R}\right\rangle \right) ^{2}\right\rangle _{c}=%
\frac{4}{N}\partial _{\beta }^{2}\ln (Z_{c})
\end{equation}

Moreover the energy density $\varepsilon (\beta )$ is given by:

\begin{equation}
\varepsilon (\beta )=\frac{1}{2\beta }-\frac{1}{N}\partial _{\beta }(\ln
Z_{c})
\end{equation}

In the following, when dealing with microcanonical estimates, 
this expression will be implicitly systematically 
used to express $\beta$ as a function of the energy density. We define also the 
two notations $\kappa _{0} \equiv \left\langle k_{R}\right\rangle _{\mu }$ and 
$\sigma _{\kappa }^{2} \equiv \left\langle \delta ^{2}K_{R}\right\rangle _{\mu}$.

\section{Analytic estimate for $\lambda _{1}$ below and above the transition}

Let us now derive the analytic estimate for $\lambda _{1}$ below 
and above the transition.
Below the critical energy, the saddle-point method gives: 
\begin{equation}
Z_{c}(\beta )\simeq (2\pi )^{N}\frac{N r^{*}}{\beta c}\exp (-N\psi (r^{*},\beta
)) \sqrt{\frac{2\pi }{N\partial _{r}^{2}\psi (r^{*},\beta )}}
\end{equation}

As the ensemble averages $\left\langle
k_{R}\right\rangle _{c}$ and $\left\langle k_{R}\right\rangle _{\mu }$
coincide in the mean-field limit, this gives:

$\left\langle k_{R}\right\rangle _{\mu }=c+\frac{2}{N}\partial _{\beta }\ln
(Z_{c}){\sim }c-2\partial _{\beta }\left(
\psi (r^{*}(\beta ),\beta )\right) $

$=c-2\frac{dr^{*}}{d\beta }\left. \partial _{r}\psi \right| _{r^{*}}-2\left.
\partial _{\beta }\psi \right| _{r^{*}}=c-2(-\frac{r^{*2}}{2\beta ^{2}c}+%
\frac{c}{2})$

That is: 
\begin{equation}
\left\langle k_{R}\right\rangle _{\mu }{%
\sim }\frac{r^{*}(\beta )^{2}}{c\beta ^{2}}
\label{courmu}
\end{equation}

Remember that $k_{R}$ is proportional to the square norm of the magnetization
(\ref{courbem}) so that we expect it to exhibit the same behavior at the transition
point with twice the characteristic exponent. Actually, a straightforward
expansion near the transition leads to

\[
\left\langle k_{R}\right\rangle _{\mu }{%
\sim }\frac{2(\beta c-2)}{\beta }=\frac{8}{1+4c}(\varepsilon_{c}-\varepsilon)
\text{  for }\varepsilon_{c}\gtrsim \varepsilon. 
\]

Taking into account the correction (\ref{correc}) and noting that $\partial _{\beta }\left\langle k_{R}\right\rangle _{c}=\frac{1}{2}%
\left\langle \delta ^{2}K_{R}\right\rangle _{c}$, one finally obtains:
 
\begin{equation}
\left\langle \delta ^{2}K_{R}\right\rangle _{\mu }=\left\langle \delta
^{2}K_{R}\right\rangle _{c}\left( 1+\frac{\beta ^{2}}{2}\left\langle \delta
^{2}K_{R}\right\rangle _{c}\right) ^{-1}
\label{flucmu}
\end{equation}

with $\left\langle \delta ^{2}K_{R}\right\rangle _{c}=\frac{4}{N}\partial _{\beta
}^{2}\ln (Z_{c}) {\sim }\frac{4r^{*}}{%
\beta ^{2}c}(\partial _{\beta }r^{*}-\frac{r^{*}}{\beta })$.

Figure 1 displays the behaviors of both the average Ricci curvature 
$\kappa _{0}$ and fluctuations $\sigma_{\kappa}$, with the control parameter 
$c$ put equal to $1$ in both figures.
Using (\ref{courmu},\ref{flucmu}), one can then 
derive $\lambda _{1}(\varepsilon )$ in the clustered phase. 
The result, obtained through (\ref{lambda},\ref{glambda},\ref{tau}), is reported 
in Fig. 2. When $\varepsilon$ approaches $\varepsilon_{c}$, expanding 
the expression for the largest Lyapunov exponent $\lambda_{1}(\varepsilon)$ 
provides the scaling law

\begin{equation}
\lambda_{1}(\varepsilon)\propto (\varepsilon_{c}-\varepsilon)^{\frac{1}{6}}
\label{critexp}
\end{equation} 
associating thereby a critical exponent, equal to $1/6$, to the dynamical 
observable $\lambda_{1}$.

Above the critical energy, one obtains in the same way:

\begin{equation}
Z_{c}(\beta )\simeq (2\pi )^{N}\exp (-N\frac{\beta c}{2})\left( 1-\frac{%
\beta c}{2}\right) ^{-1}
\end{equation}

Here, as $\left\|{\bf M}\right\|^{2}$ becomes of order 
$O\left( N^{-1}\right)$, we shall use 
the full expression 
$k_{R}=c\left\|{\bf M}\right\|^{2}-\frac{c}{N} + O\left( N^{-2}\right)$. 
Then: 
\begin{equation}
\left\langle k_{R}\right\rangle _{\mu }=
\frac{\beta c^{2}}{N(2- \beta c)}+O\left(
N^{-2}\right)
\end{equation}

i.e. the microcanonical average of the Ricci curvature 
vanishes in the mean-field
limit.

Similarly: 
\[
\left\langle \delta ^{2}K_{R}\right\rangle _{c} = \frac{4}{N}\partial _{\beta
}^{2}\ln (Z_{c})=\frac{4 c^{2}}{N}\left( 2- \beta c \right)
^{-2}=O\left( N^{-1}\right) 
\]

As $\varepsilon (\beta )\sim \frac{1}{2\beta }+\frac{c}{2}$, the correcting
term needed to get the microcanonical fluctuations is of order $N^{-2}$,
thus negligible. And then:

\begin{equation}
\left\langle \delta ^{2}K_{R}\right\rangle _{\mu } \sim \frac{4 c^{2}}{N}\left( 2-%
\beta c \right) ^{-2}=O\left( N^{-1}\right)
\end{equation}

We can keep in further calculations the dominant order in $N$, and derive
the scaling law with $N$ for the largest Lyapunov exponent. Using expressions 
(\ref{lambda},\ref{glambda},\ref{tau}) 
, in the limit $N\rightarrow \infty $, one obtains

\begin{equation}
\lambda _{1}{\sim }%
{\frac{ 4^{\frac{1}{3}} c \sqrt{ \beta c}   }%
  { (2- \beta c) ^{\frac{3}{2} }  } }%
N^{-\frac{1}{3}}
\label{scaling}
\end{equation}

\section{Comments and conclusions}
Let us first comment here on the reliability expected for the expressions 
just derived. As developped in refs. \cite{Casetti95,Casetti96,Caiani97,Pettini}, the geometrical approach aims at extracting information 
on, at least, an average degree of chaoticity of the dynamics 
from mean global geometrical properties of the Riemannian manifold 
constructed from a given Hamiltonian.
This implies the crucial assumption of ergodicity, as a way of bypassing the 
knowledge of the trajectories, i.e. the numerical integration of the 
equations of motion. This ergodic hypothesis is not expected to be realized 
in the integrable limits of small and large energy, the latter following 
from the boundedness of the potential energy in (\ref{H}). However, it is 
well known that chaos is not a necessary condition for ergodicity, the most 
striking piece of evidence being provided by the ideal gas of point particles, 
for which there is no velocity mixing at all. Also 
recent studies \cite{Livi} have emphasized that ergodic-like properties 
should depend mainly on the observable at hand, irrespectively of the 
degree of chaoticity of the dynamics. 
Concerning our model, S. Ruffo already observed in 
\cite{Ruffo94} a good agreement between Gibbs predictions and numerical 
simulations for the observable $\left\|{\bf M}\right\|$. Moreover, in the 
mean-field limit, this happens even in the 
integrable limit of large energy, an explanation for this being provided 
by a result of M. Kac \cite{Ruffo94,Kac}, so that the mean-field magnetization 
appears like a good observable with respect to ergodicity.
 Therefore it is not surprising to observe that numerical calculations of 
the mean Ricci curvature and its variance fit well the microcanonical 
predictions presented in Fig. 1 \cite{Latora97,Rapi}, except in the vicinity of the 
phase transition where finite-$N$ effects dominate.  
Concerning the transition region, 
as noted before, the formula \cite{Lebo} used to get fluctuations in the 
microcanonical ensemble from canonical ones is not valid at the 
critical energy. Therefore we should exclude in our conclusions a small 
neighborhood of $\varepsilon _{c}$, all the smaller as $N$ is large. So the 
analytic estimate for $\lambda_{1}(\varepsilon )$ in the mean-field limit 
is expected to be quite reliable except maybe for small $\varepsilon$ and in 
the vicinity of $\varepsilon _{c}$. It should also be noted that the timescale 
$\tau$ estimated as (\ref{tau}), that is the time under which the effective 
curvature felt by a geodesic cannot be regarded as a random process, is 
the less solid point of the geometrical modeling \cite{Casetti96,Caiani97} as 
(\ref{tau}) relies mainly on phenomenological arguments. Then it can, if 
necessary, be slightly adjusted to fit numerical calculations.
Nonetheless that estimate for $\tau$ is also a powerful tool, as it provides 
a natural timescale, depending on $\varepsilon$, 
that should be taken into account 
to connect for instance results for mappings \cite{Paladin86} 
to results for continuous flows as it is the case here.\\
Keeping these remarks in mind, we can now comment on the results obtained 
in Sec. III.
Expression (\ref{scaling}) means that, in this mean-field model, above the 
critical energy, chaos does not survive to the limit 
$N\longrightarrow \infty $. 
This can be conjectured straightforwardly from the equation (\ref{pendule}) 
governing the time evolution of any particle, which predicts ballistic motion 
as $\left\|{\bf M}\right\|$ vanishes above $\varepsilon _{c}$.
Moreover, one obtains the scaling law $N^{-\frac{1}{3}}$ for the largest Lyapunov
exponent to the leading order in $N$. The same scaling law has been found
numerically by Latora et al. \cite{Latora97}. A rather nice
fit (see Fig. 2) is also obtained with Yamaguchi's simulations \cite{Yama96} 
on a wide range of $\varepsilon$, except in the vicinity of $\varepsilon _{c}$, where finite size effects smooth the
transition. Here strong metastability related to critical slowing down 
may also affect numerical results with 
relaxation times towards equilibrium increasing greatly with $N$. Besides, for a 
given $N$ large enough, expression (\ref{scaling}) rightly gives a 
vanishing Lyapunov exponent in the integrable limit of large energy where 
rotators tend to behave as free particles.\\
Concerning the transition region, in spite of the above mentioned 
remarks on the validity of our results at the 
critical energy, let us mention the remarkable features exhibited by 
Figs. 1,2: $\kappa _{0}$, $\sigma _{\kappa }$
and $\lambda _{1}$ display singular behaviors at the critical point.
Here curvature fluctuations exhibit a discontinuity which is similar to 
the ``cusp'' numerically observed in \cite{Caiani97}. In our case, this 
appears as a direct consequence of the second order phase transition 
exhibited by the model and, following equation (\ref{courbem}), 
of the expressions of the different parameters used 
in the geometrical approach in terms of smooth functions of the order parameter. 
Following conjectures exposed in \cite{Caiani97},
the geometrical meaning of these singular behaviors might be that a 
topology change of the ``mechanical'' manifold underlying the dynamics 
occurs at the critical energy.\\
Finally, as for $\lambda _{1}$, its maximal value would be reached slightly 
below the critical point and not at the critical point. 
Numerical simulations made in \cite{Latora97} 
for 20000 particles show such a tendency. Moreover, when $\varepsilon$ 
approaches the critical energy, 
calculations (\ref{critexp}) show that $\lambda _{1}$ goes to $0$ 
as $(\varepsilon_{c}-\varepsilon)^{\frac{1}{6}}$. This suggests that a 
critical exponent could be associated to the largest Lyapunov 
exponent as a dynamical observable.\\
Further studies 
should inspect more precisely the region where the amplitudes 
of the curvature and fluctuations are comparable, around $\varepsilon=0.45$ 
(see Fig. 1). As observed in other models, 
for such a situation strong stochasticity may be expected. 
A more refined treatment may imply some corrections to the 
gaussianity of the effective curvature, that would take into 
account further moments of the mean Ricci curvature. Also the vicinity of the 
critical energy, as well as a possible extension of the results 
obtained in this article to a larger class 
of mean-field Hamiltonian systems deserve obviously further investigations.

% Acknowledgments

\acknowledgments

The author is greatly indebted to Y. Elskens and M. Pettini for their 
advice and explanations, and thanks M. Antoni and S. Ruffo for fruitful 
communications. MCF is supported by a grant from the Minist\`{e}re de 
l'enseignement sup\'{e}rieur et de la recherche. This work is part of the 
European research network on stability and universality in classical 
mechanics (contract ERBCHRXCT940460).

%% figure captions

\clearpage

\noindent
{\bf Figure captions}

\vskip 1 cm
{\bf Fig. 1:}
Analytic expressions for the microcanonical averages of the 
average Ricci curvature, $\kappa_{0}$ (solid curve) and of its fluctuations 
$\sigma _{\kappa }$ (dot-dashed curve) in the mean-field limit, 
below and above the phase transition.

\vskip 1 cm
{\bf Fig. 2:}
Analytic expression for the largest Lyapunov exponent $\lambda_{1}$ 
in the mean-field limit (solid curve) below and above the phase transition. 
Analytic corrections (dot-dashed curves) to mean-field limit 
for finite $N$ with $N=80$ and $N=200$ above $\varepsilon_{c}$.
Here the derivation does not restrict to the leading term (\ref{scaling}) but 
computes (\ref{lambda},\ref{glambda},\ref{tau}) up to further orders, as $N$ is 
not very large. There is a nice fit with results exposed in \cite{Yama96} 
apart from the vicinity of the critical energy.


\begin{thebibliography} {99}

\bibitem{Paladin86}  G. Parisi and A. Vulpiani, J. Phys. A\ {\bf 19}, L425\ (1986).

\bibitem{Casetti95}  L. Casetti, R. Livi and M. Pettini, Phys. Rev. Lett. {\bf 74},
375\ (1995).

\bibitem{Casetti96}  L. Casetti, C. Clementi and M. Pettini, Phys. Rev. E {\bf 54},
5469 (1996).

\bibitem{Caiani97}  L. Caiani, L. Casetti, C. Clementi and M. Pettini, 
LANL preprint chao-dyn/9702011; L. Caiani, L. Casetti, C. Clementi, 
G. Pettini, M. Pettini and R. Gatto, LANL preprint hep-th/9706081.

\bibitem{Pettini} M. Pettini, Phys. Rev. {\bf 54}, 828 (1993) and references 
quoted therein.

\bibitem{Szasz}  D. Sz\'asz, ``Boltzmann's Ergodic Hypothesis, a Conjecture
for Centuries? '', Lecture given at {\it The international symposium in
honour of Boltzmann's 150}$^{th}${\it \ birthday}, Vienna, Feb. 24-26, 1994,
Preprint ESI 98 (1994).

\bibitem{Cerruti}  M. Cerruti-Sola and M. Pettini, Phys. Rev. E {\bf 53},
179 (1995).

\bibitem{Kandrup}  H. E. Kandrup, Phys. Rev. E {\bf 56}, 2722 (1997).

\bibitem{Latora97}  V. Latora, A. Rapisarda and S. Ruffo, Phys. Rev. 
Lett. {\bf 80}, 692 (1998).


\bibitem{Yama96}  Y. Y. Yamaguchi, Progr. of Theor. Phys. {\bf 95},
717 (1996).

\bibitem{Antoni95}  M. Antoni and S. Ruffo, Phys. Rev. E {\bf 52}, 2361 (1995).\ 

\bibitem{Ruffo94}  S. Ruffo, in {\it Transport and Plasma Physics}, edited
by S. Benkadda, Y. Elskens and F. Doveil (World Scientific, Singapore, 1994), 
p.  114-119.

\bibitem{Elskens97}  Y. Elskens and M. Antoni, Phys. Rev. E {\bf 55}, 6575 (1997).

\bibitem{Anto} The equivalence of canonical and microcanonical ensembles in
 the mean-field limit for this model has recently been explicitly proved, 
 M. Antoni (private communication).
 
\bibitem{Balian}  R. Balian, {\it From Microphysics to Macrophysics - 
Methods and Applications of Statistical Physics} (Springer-Verlag, Berlin, 
1991).

\bibitem{Lebo}  J. L. Lebowitz, J. K. Percus and L. Verlet, Phys.
Rev. {\bf 153}, 250 (1967).

\bibitem{note1} A rigourous definition would replace $N$ by $N-1$, 
but this would contribute to negligible terms throughout the paper.

\bibitem{Livi} C. Giardin\'a, R. Livi, LANL preprint chao-dyn/9709015.

\bibitem{Kac} M. Kac, Am. J. Math. {\bf 65}, 609 (1943).

\bibitem{Rapi} V. Latora, A. Rapisarda and S. Ruffo, unpublished.

\end{thebibliography}
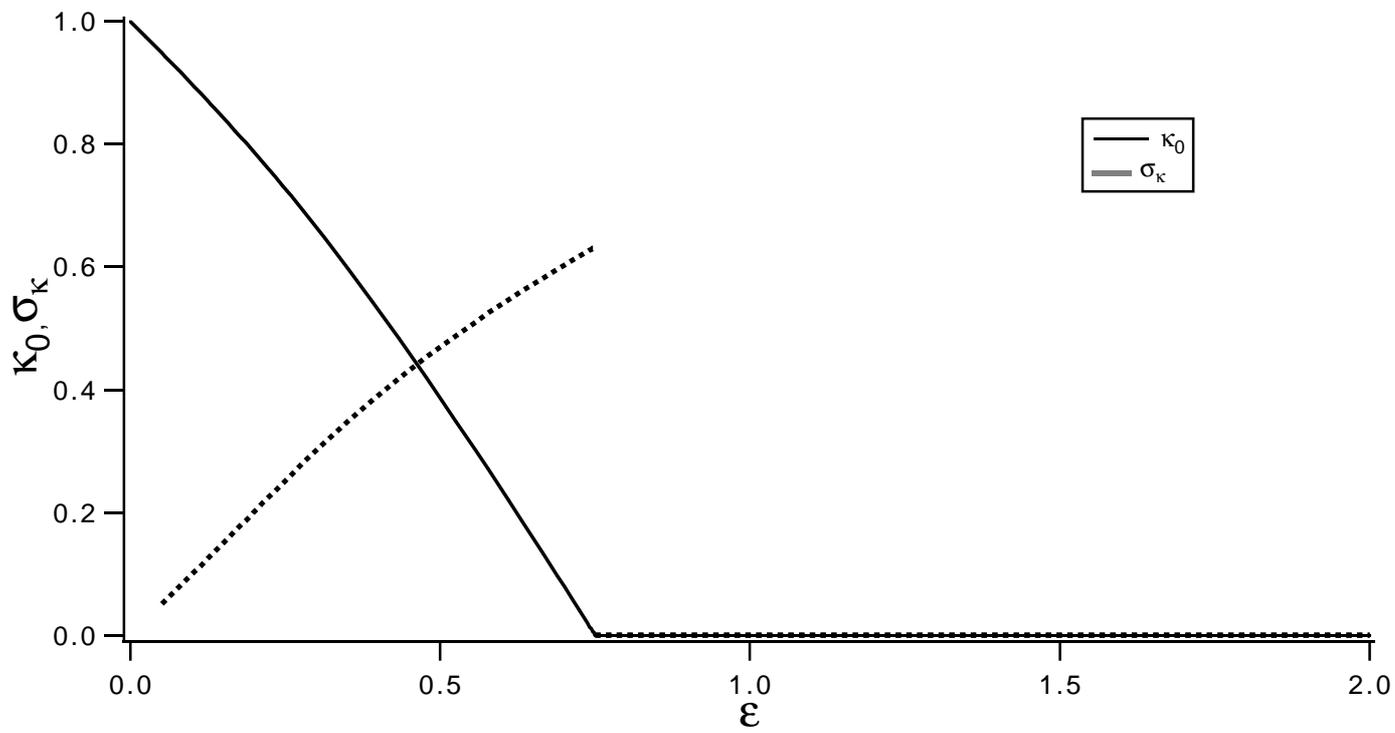
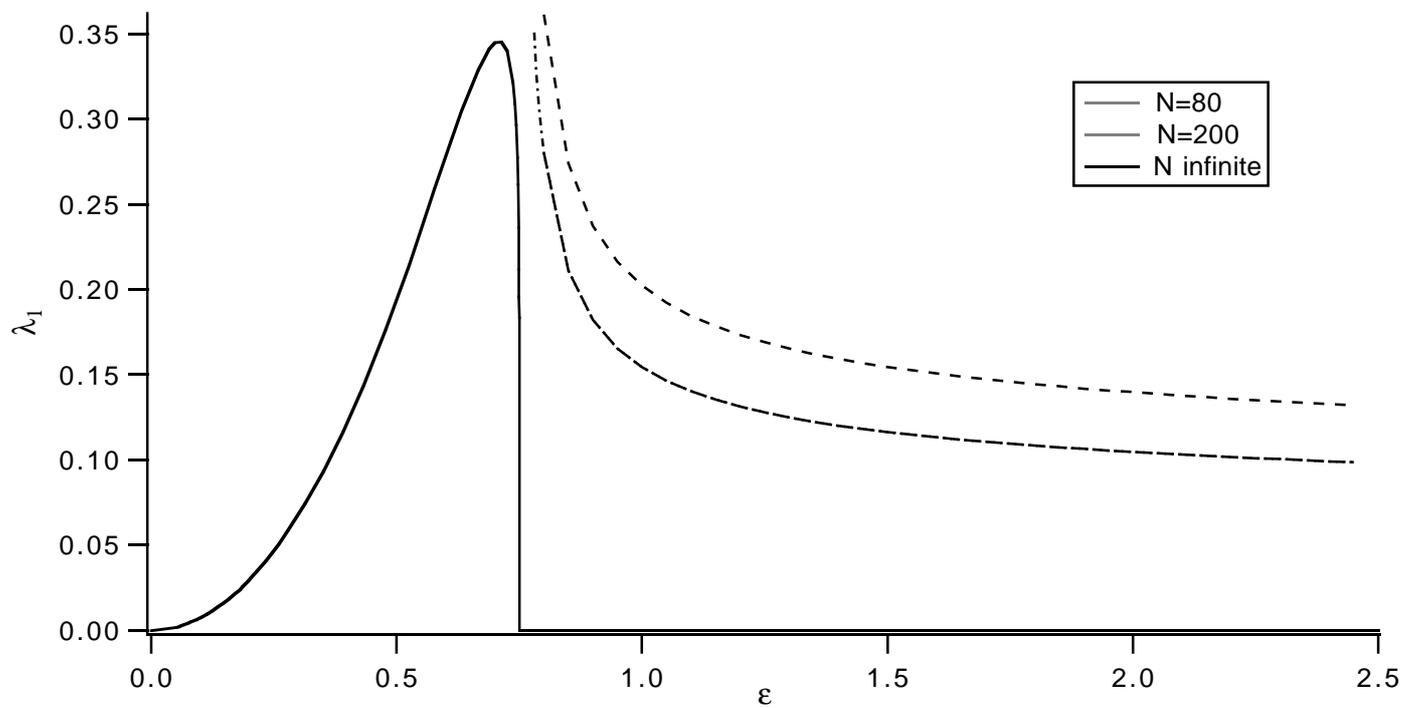
\end{document}